\documentclass[twocolumn,pra,superscriptaddress]{revtex4-2}

\usepackage{epsfig}
\usepackage{graphicx}
\usepackage{amsmath}
\usepackage{amssymb}
\usepackage{epstopdf}
\usepackage{xcolor}
\usepackage{bm}
\usepackage{dsfont}
\usepackage{hyperref}
\usepackage{braket}
\usepackage{tikz}
\usepackage[normalem]{ulem}
\usepackage{subcaption}

\begin{document}
	
	\title{Quantum optical formulation of difference-frequency generation and optimal cloning of spatial modes}
	
	\author{Tanita Permaul}
	\affiliation{School of Chemistry and Physics, University of KwaZulu-Natal, Durban, South Africa}
	\author{Andrew Forbes}
	\affiliation{School of Physics, University of Witwatersrand, Wits, South Africa}
	\author{Thomas Konrad}
	\email{konradt@ukzn.ac.za}
	\affiliation{School of Chemistry and Physics, University of KwaZulu-Natal, Durban, South Africa}
	\affiliation{National Institute of Theoretical and Computational Sciences (NITheCS), KwaZulu-Natal, South Africa}
	
	\begin{abstract}
		We present a quantum optical formulation of difference-frequency generation (DFG) that incorporates the spatial modes of light. It reproduces the well established result for classical light beams and establishes the relation of DFG to stimulated and spontaneous parametric downconversion. As shown here, these relations determine that stimulated parametric down-conversion can realise $N\xrightarrow{} M $ $d$-dimensional optimal quantum cloning.     
	\end{abstract}

	\maketitle
	
	\section{Introduction}
	Nonlinear optics is experiencing a resurgence of late, as novel materials \cite{xu2022femtosecond,guo2023ultrathin} and the nonlinear interaction with novel forms of light \cite{buono2022nonlinear} extend our understanding beyond the textbook. This has resulted in new selection rules \cite{wu2023observation,chen2020phase,tang2020harmonic} and processes \cite{wu2022conformal,luttmann2023nonlinear,da2022observation,da2021stimulated} based on structured light \cite{forbes2021structured}, spatial mode control beyond frequency \cite{sephton2019spatial,singh2024light} and has recently been deployed in high-dimensional teleportation with spatial modes of light \cite{sephton2023quantum, qiu2023remote}. Despite these exciting advances, no account as yet exists that marries a classical and quantum picture of nonlinear processes with spatially structured photons. The process of difference-frequency generation (DFG) caused by the interaction of light with a dielectric medium is well established in classical nonlinear optics for producing a beam with the desired frequency at the difference-frequency of two input beams. Due to energy conservation, this process also amplifies the input beam with the lower frequency. This amplification feature of the DFG process is normally referred to as optical parametric amplification, or optical parametric oscillation when the nonlinear crystal is placed in a cavity \cite{boyd}. DFG usually involves coherent (classical) states of light, suggesting a classical description by means of Maxwell's equations for electromagnetic fields. However, at the same time, the nonlinear response of the medium also generates spontaneous parametric down-conversion (SPDC), a competing process, that can produce entangled photon pairs, which can only be appropriately represented in quantum theory. 
	
	Hence, a complete picture of DFG requires a  quantum description. At the quantum level, DFG  corresponds to stimulated parametric down-conversion (StimPDC), a quantum effect, that is also shown to achieve optimal cloning of polarisation qubits \cite{simon2000optimal}, and of time-bin qudits \cite{cerf2006optical}. However, due to limitations of the current description of StimPDC \cite{oliveira2019phase}, it is difficult to extend this to higher dimensions for the spatial degrees of freedom of light. In this paper, we present a quantum optical description of DFG, i.e. StimPDC, in terms of spatial modes. We arrive there by considering the formulation of SPDC with a more general mode expansion than the usual Schmidt decomposition. 
	
	An application of the newly derived quantum description for the output spatial modes of StimPDC is to consider optimal quantum cloning of spatial modes of light. As is well known, it is not possible to create perfect copies of an unknown quantum state \cite{wootters1982single}. This is referred to as the \textit{no-cloning theorem}, which is fundamental to secure quantum communication schemes. However in 1996, Bu\v{z}ek and Hillery \cite{buvzek1996quantum} demonstrated that quantum cloning can be achieved partially with imperfect clones. Optimal quantum cloning produces clones with the highest fidelities allowed by quantum mechanics. A quantum cloning device is described as universal when all input states are cloned equally well, and symmetric when the fidelities of all of the clones are the same. Optimal cloning of orbital angular momentum (OAM) qubits \cite{nagali2009optimal} and ququarts \cite{nagali2010experimental} has been demonstrated via symmetrization with the Hong-Ou-Mandel effect. These schemes only work on the single photon level with limited spatial modes, but a potential scheme utilising StimPDC would theoretically be able to employ many photons and an unlimited number of spatial modes. The optimality of such a cloning scheme is proved below.
	
	\section{StimPDC}
	In this section, we present the quantum optical formulation of DFG that incorporates spatial modes. The phenomenon of SPDC is a quantum nonlinear effect with no classical analog. However, the process of StimPDC does have a classical limit given by the second-order nonlinear optical process of DFG. We derive a quantum optical formulation for StimPDC by modifying the model for SPDC \cite{walborn2010spatial} to incorporate the presence of a seed beam. Considering a general mode decomposition instead of a Schmidt decomposition of the output mode light, allows us to recover the product rule of DFG, which states that the output mode is given in terms of a product of the complex conjugate of one of the input modes with the other input mode.
	
\begin{figure}
		\includegraphics[width=\linewidth]{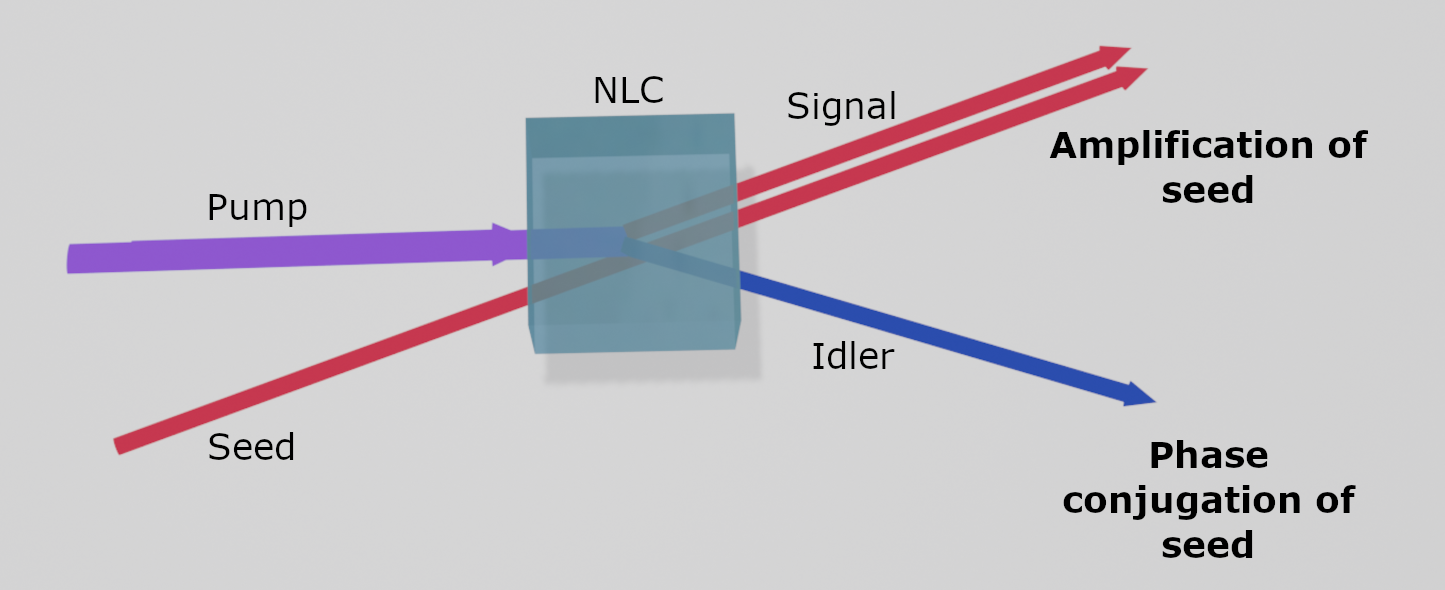}
		\caption{\label{fig:dfg} Difference-frequency generation schematic. Photons from a seed beam stimulate parametric down conversion of pump photons into the signal (same spatial light mode and frequency as seed beam), and idler with complex conjugate spatial mode and difference frequency.}
	\end{figure}

	As DFG is a second-order nonlinear process, it involves the interaction of three light fields, which are usually denoted as the pump field with frequency $\omega_1$, the signal field with frequency $\omega_2$ and the idler field with frequency $\omega_3=\omega_1-\omega_2$, as shown in Fig.~\ref{fig:dfg}. If we consider the process on the single photon level; the signal photon from a seed beam stimulates the down-conversion of the pump photon such that an idler photon is produced at the difference-frequency of the two input photons.
	
	The nonlinear interaction Hamiltonian $H_I$ for parametric down-conversion is due to the interaction of the incoming electric field $\textbf{E}$ and the nonlinear polarization $\textbf{P}$ of the medium:
	\begin{equation}
		\begin{split}
			H_I& = \frac{1}{2}\int_V \textbf{E} \cdot \textbf{P} \mathrm{d}^3\textbf{r} \\
			&= -i  \chi^{(2)} \frac{\hbar^{3/2}\sqrt{\omega_1 \omega_2 \omega_3}}{\sqrt{\varepsilon_0}(2 \pi )^3} \int_V \mathrm{d}^3\textbf{r} \int_V \mathrm{d}^3\textbf{k}_1  \mathrm{d}^3\textbf{k}_2 \mathrm{d}^3\textbf{k}_3  \\
			& \mathrm{e}^{-i(\textbf{k}_3+\textbf{k}_2-\textbf{k}_1)\cdot \textbf{r}} \mathrm{e}^{i(\omega_3+\omega_2-\omega_1)t}
			\hat{a}(\textbf{k}_1) \hat{a}^{\dagger}(\textbf{k}_2) \hat{a}^{\dagger}(\textbf{k}_3) + \mbox{h.c.}\ .
		\end{split}
	\end{equation}
	Here $\chi^{(2)}$ is the second order nonlinear susceptibility of the crystal, $\hbar$ is the reduced Planck's constant, $\varepsilon_0$ is the permittivity of free space, $\textbf{k}$ is the wave vector for a particular field, and $\hat{a}^{(\dagger)}$ denotes an annihilation (creation) operator.  This is simplified by evaluating the time and volume integrals which reveal the phase and frequency-matching conditions of the process:
	\begin{equation}
		\begin{split}
			&\int\mathrm{e}^{i(\omega_3+\omega_2-\omega_1)t} \mathrm{d}t = 2\pi\ \delta(\omega_1-\omega_2-\omega_3) ,\\
			&\int_A  \mathrm{e}^{i(\textbf{q}_{1}-\textbf{q}_{2}-\textbf{q}_{3})\cdot \textbf{x}} \mathrm{d}^2\textbf{x}= (2\pi)^2\delta^{2}(\textbf{q}_{1}-\textbf{q}_{2}-\textbf{q}_{3}) ,\\
			&\int_0^L  \mathrm{e}^{i(k_{1z}-k_{2z}-k_{3z})z} \mathrm{d}z\approx \mathrm{sinc}\left(\frac{\Delta k_z L}{2}\right).\\
		\end{split}
	\end{equation}
	We can perform these integrations by assuming that the interaction time is long enough and that the transverse profile of the beam is much smaller than the interaction surface area. The integration over the finite length of the crystal $L$ results in a sinc function that depends on the phase mismatch $\Delta k_z = k_{1z}-k_{2z}-k_{3z}$.
	The unitary operator can therefore be expressed as 
	\begin{equation}\label{unitary}
		\begin{split}
			\hat{U}(t)&=\text{exp}\left(\frac{1}{i\hbar}\int_0^t \text{d}\tau \hat{H}_I(\tau)\right)\\
			&= \text{exp}\left(C\int \mathrm{d}^3\textbf{k}_1 \mathrm{d}^3\textbf{k}_2 \mathrm{d}^3\textbf{k}_3\ \delta^2(\textbf{q}_{1}-\textbf{q}_{2}-\textbf{q}_{3})\ \times \right. \\
			&\left. \ \times\, \mathrm{sinc}\left(\frac{\Delta k_z L}{2}\right)\hat{a}(\textbf{k}_1)\hat{a}^{\dagger}(\textbf{k}_2)  \hat{a}^{\dagger}(\textbf{k}_3) +\mbox{h.c.}\right).\\
		\end{split}
	\end{equation} 
	The total output state of the down-conversion process is given by $\ket{\psi_{\text{T}}}=\hat{U}\ket{\psi_0}$. For a sufficiently weak pump beam, the unitary operator can be expressed by a power series expansion: $\hat{U}(t)= 1+ \left(\frac{1}{i\hbar}\int_0^t \text{d}\tau \hat{H}_I(\tau) \right)+ \dots\ $. Higher order terms in the expansion create multiple down-converted pairs, based on the number of pump photons that are down-converted. We now show how to obtain the biphoton state due to a single pump photon being down-converted, however, this can easily be extended to the case of higher order down-conversions. The output biphoton state, from the initial state $ \ket{\psi_0}=\ket{1_{U_{p}}}\ket{1_{S_{l}}}$, is given by
	\begin{equation}\label{eqn:psi2full}
		\begin{split}
			&\ket{\psi_2}=\left(C\int \mathrm{d}^3\textbf{k}_1 \mathrm{d}^3\textbf{k}_2 \mathrm{d}^3\textbf{k}_3\ \delta^2(\textbf{q}_{1}-\textbf{q}_{2}-\textbf{q}_{3})\  \right.\\
			&\left. \times\, \mathrm{sinc}\left(\frac{\Delta k_z L}{2}\right) \mathcal{U}_p (\textbf{q}_1)\hat{a}^{\dagger}(\textbf{k}_2)  \hat{a}^{\dagger}(\textbf{k}_3)\right)\ket{0}_p\ket{1_{S_{l}}}_s \ket{0}_i ,\\
		\end{split}
	\end{equation} 
	where $\mathcal{U}_p (\textbf{q}_1)$ is the spatial mode of the pump beam at the front plane of the crystal and $\mathcal{S}_{l}(\textbf{q}_2)$ is the spatial mode of the initial signal (seed) photon. The spatial mode functions are here given in their spectral representation as functions of the transversal spatial frequencies $\textbf{q}$. 
	The component due to the Hermitian conjugate (h.c.) term from Eq.~(\ref{unitary}) vanishes due to $\hat{a}(\textbf{k}_3)\ket{0}_i=0$. We define the mode function 
	\begin{equation}\label{mode_eq}
		\Phi(\textbf{q}_2,\textbf{q}_3)=\mathcal{U}_p (\textbf{q}_2+\textbf{q}_3)\, \mathrm{sinc}\left(\frac{\Delta k_z L}{2}\right),
	\end{equation}
	where the sinc function is the longitudinal phase-matching function. We can separate the wave vector, $\textbf{k}=\textbf{q}+k_z \textbf{u}_z$, into its transverse and longitudinal components. Thereafter Eq.~(\ref{eqn:psi2full}) reads
	\begin{equation}\label{eqn:psi2}
		\begin{split}
			\ket{\psi_2}&=\left(C\int \mathrm{d}^3k_{1z} \mathrm{d}^3k_{2z} \mathrm{d}^3k_{3z} \mathrm{d}^2\textbf{q}_2 \mathrm{d}^2\textbf{q}_3\ \times \right.\\
			&\times \left. \Phi(\textbf{q}_2,\textbf{q}_3) \hat{a}^{\dagger}(\textbf{k}_2)  \hat{a}^{\dagger}(\textbf{k}_3)\right)\ket{0}_p\ket{1_{S_{l}}}_s \ket{0}_i ,\\
		\end{split}
	\end{equation} 
	where the integral over $\textbf{q}_1$ has been evaluated. Due to the paraxial nature of the interacting light, it is sufficient to focus on just the transverse component.
	Therefore, for a nonlinear crystal pumped along the z-direction with monochromatic and paraxial approximations, the transverse profile of the biphoton down-conversion output state with an initial signal photon is given by \cite{ribeiro1999image},
	\begin{equation}\label{output}
		\ket{\psi_{\text{out}}}= C \int \mathrm{d}^2\textbf{q}_2 \mathrm{d}^2\textbf{q}_3 \Phi(\textbf{q}_2,\textbf{q}_3) \hat{a}^{\dagger}(\textbf{q}_2) \ket{1_{S_{l}}}_s \hat{a}^{\dagger}(\textbf{q}_3)\ket{0}_i,
	\end{equation}
	where constant $C = (\hbar \sqrt{\omega_1 \omega_2 \omega_3}\ \mathcal{\chi}_{\text{eff}}^{(2)})/\sqrt{\varepsilon_0}$ contains the coefficients due to the nonlinear process, $\textbf{q}_{2,3}$ are the transverse wave-vector components for the signal ($s$) and idler ($i$) photons respectively, and $\ket{1_{S_{l}}}_s$ is the initial signal photon:
	\begin{equation}\label{insig}
		\ket{1_{S_{l}}}=\int \mathrm{d}^2\textbf{q}_2 \: \mathcal{S}_{l}(\textbf{q}_2) \hat{a}^{\dagger}(\textbf{q}_2) \ket{0}_s.
	\end{equation}
	 The mode function $\Phi(\textbf{q}_2,\textbf{q}_3)$ can be decomposed in terms of the idler and signal modes as
	\begin{equation}
		\Phi(\textbf{q}_2,\textbf{q}_3)=\sum_{k,m} \mathcal{N}^p_{k,m} \:\mathcal{I}_{k}(\textbf{q}_3) \mathcal{S}_{m}(\textbf{q}_2) , 
	\end{equation}
	where $\mathcal{I}_{k}(\textbf{q}_3)$ and  $\mathcal{S}_{m}(\textbf{q}_2)$ are complete sets of orthonormal spatial modes for the idler and signal modes, respectively. In particular, the basis set of signal modes contains for $m=l$ the initial (seed) mode $\mathcal{S}_l$. This decomposition is not necessarily a Schmidt decomposition. The Schmidt decomposition is a single sum expansion which is often used to describe SPDC with a Gaussian pump beam. Here we chose the more general decomposition into a product basis which involves independent summation over the basis modes for the idler and signal beams. 
	
	This is similar to writing a composite state $\ket{\Psi}_{1,2} \in \mathcal{H}_1 \otimes \mathcal{H}_2$ in terms of the basis states $(\ket{\varphi_k})_k$,  $(\ket{\psi_m})_m$ of the individual systems, $\ket{\Psi}_{1,2}=\sum_{k,m} c_{k,m} \ket{\varphi_{k}}_1 \ket{\psi_{m}}_2 $. The wave function is therefore $\Psi(\textbf{r}_1,\textbf{r}_2)= \braket{\textbf{r}_1,\textbf{r}_2|\Psi}_{1,2}=\sum_{k,m} c_{k,m} \braket{\textbf{r}_1|\varphi_{k}}_1 \braket{\textbf{r}_2|\psi_{m}}_2 =\sum_{k,m} c_{k,m} \varphi_{k}(\textbf{r}_1) \psi_{m}(\textbf{r}_2).$  
	
	The product basis expansion coefficients $ \mathcal{N}^p_{k,m}$ are calculated as inner products corresponding to integrals over the variables:
	\begin{equation}\label{deq:dfgco}
		\mathcal{N}^p_{k,m}=\int \mathrm{d}^2\textbf{q}_2 \mathrm{d}^2\textbf{q}_3 \: \mathcal{I}_{k}^*(\textbf{q}_3) \mathcal{S}_{m}^*(\textbf{q}_2) \Phi(\textbf{q}_2,\textbf{q}_3) .
	\end{equation}
	Using this mode function expansion, we can now express the output state as
	\begin{equation}
		\begin{split}
			\ket{\psi_{\text{out}}}=\sum_{k,m} \int & \mathrm{d}^2\textbf{q}_2 \mathrm{d}^2\textbf{q}_3 \mathcal{N}^p_{k,m}\: \mathcal{I}_{k}(\textbf{q}_3)
			\mathcal{S}_{m}(\textbf{q}_2) \\
			& \hat{a}^{\dagger}(\textbf{q}_2) \ket{1_{S_{l}}}_s\hat{a}^{\dagger}(\textbf{q}_3)\ket{0}_i. \\
		\end{split}
	\end{equation}
	Defining operators that create a photon in a specific spatial mode (e.g. $\mathcal{S}_m(\textbf{q}_j)$) as
	\begin{equation}
		\hat{a}^{\dagger}_{S_m}=\int \mathrm{d}^2\textbf{q}_j \mathcal{S}_m(\textbf{q}_j) \hat{a}^{\dagger}(\textbf{q}_j),
	\end{equation} 
	we can simplify the output state to be,
	\begin{equation}\label{ddout}
		\begin{split}
			\ket{\psi_{\text{out}}}&=\sum_{k}  \mathcal{N}^p_{k,l}\sqrt{2} \ket{2_{S_{l}}}_s \ket{1_{I_k}}_i \\
			&+\sum_{m\neq l}\sum_{k}  \mathcal{N}^p_{k,m} \ket{1_{S_{l}},1_{S_{m}}}_s \ket{1_{I_k}}_i,
		\end{split}
	\end{equation}
	where we omitted an overall normalisation constant. 
	This expression shows that the output of parametric down-conversion is made up of two parts. The first term in Eq.~(\ref{ddout}) represents stimulated down-conversion with a Bose enhancement factor of $\sqrt{2}$ due to photon-statistics. It corresponds to DFG and parametric amplification at the single-photon level.   The second term in Eq.~(\ref{ddout}) is due to spontaneous down-conversion, which occurs independently from seed photons in the signal beam. However, Bose enhancement increases with the number $n$ of initial photons in the signal beam according to $\hat{a}^{\dagger}_{S_l}\ket{n_{S_{l}}} = \sqrt{n+1}\ket{n_{S_{l}}+1}$. In the classical limit $n\xrightarrow[]{}\infty$, StimPDC thus dominates and SPDC becomes negligible. This corresponds to classical difference-frequency generation.
	
	We now focus on the StimPDC state $\ket{\psi_d}$ (first term Eq.~(\ref{ddout})) only,
	\begin{equation}
		\ket{\psi_d}\equiv \sqrt{2} \ket{2_{S_{l}}}_s \ket{1_{D_l}}_i
	\end{equation}
	with 
	\begin{equation}\label{eqn:prodidler}
		\ket{1_{D_l}}_i=\sum_{k} \mathcal{N}^p_{k,l}\ket{1_{I_k}}_i,
	\end{equation}
	and note that the idler and signal photons are in a product state in contrast to the SPDC state (second term in Eq.~(\ref{ddout})), which is entangled. 
	
	In the thin crystal approximation, the sinc in the mode function (in Eq.~(\ref{mode_eq})) is replaced by unity. As a result, the mode function equals the spatial mode of the pump beam, $\Phi(\textbf{q}_2,\textbf{q}_3)=\mathcal{U}_p (\textbf{q}_2+\textbf{q}_3)$. The Fourier integral representation of the pump mode function reads
	\begin{equation}
		\mathcal{U}_p (\textbf{q}_2+\textbf{q}_3)=\frac{1}{(2\pi)^2}\int \mathrm{d}^2\boldsymbol{\rho} \: \mathrm{U}_p(\boldsymbol{\rho}) e^{-i(\textbf{q}_2+\textbf{q}_3)\cdot \boldsymbol{\rho}}.
	\end{equation}
	This expression can be substituted into the calculation for the expansion coefficients in Eq.~(\ref{deq:dfgco}), and the integrals over $\textbf{q}_2$ and $\textbf{q}_3$ are carried out. As a result, we obtain,
	\begin{equation}\label{dcoef}
		\begin{split}
			\mathcal{N}^p_{k,l}&=\frac{1}{(2\pi)^2} \int \mathrm{d}^2\boldsymbol{\rho}\: \mathrm{I}_{k}^*(\boldsymbol{\rho}) \mathrm{S}_{l}^*(\boldsymbol{\rho}) \mathrm{U}_p(\boldsymbol{\rho})   . \\
		\end{split}
	\end{equation}
	Note, that the Fourier transform effectively diagonalises the integration kernel $\Phi(\textbf{q}_2,\textbf{q}_3)\equiv\mathcal{U}_p (\textbf{q}_2+\textbf{q}_3)$, which allows us to interpret the bilinear form (\ref{deq:dfgco}) as an inner product of the function $\mathrm{I}_{k}(\boldsymbol{\rho})$ with the function  $\mathrm{S}_{l}^*(\boldsymbol{\rho}) \mathrm{U}_p(\boldsymbol{\rho})$.
	On the other hand, $ \mathcal{N}^p_{k,l}=\braket{1_{I_k}|1_{D_l}}$, so that we can identify $\mathrm{S}_{l}^*(\boldsymbol{\rho}) \mathrm{U}_p(\boldsymbol{\rho})$ as the mode of the idler photon, i.e., 
	\begin{equation}\label{deq:idler}
		\mathrm{D}_{l}(\boldsymbol{\rho})= \mathrm{S}_{l}^*(\boldsymbol{\rho}) \mathrm{U}_p(\boldsymbol{\rho}).
	\end{equation}
	This product structure is well known from the classical treatment of DFG \cite{boyd}. Above, it is derived in the thin crystal limit for the case of a single pump photon that is down-converted in the presence of a single seed photon. But the derivation can be generalised to the case of multiple photon down-conversions  by considering higher-order expansions of the unitary operator.
	\section{Optimal Cloning}
	We now want to demonstrate that $d$-dimensional optimal quantum cloning can be achieved by StimPDC of light modes. The usual measure used for optimal quantum cloners is the single-copy fidelity \cite{gisin1997optimal}. 
	This is the fidelity of one of the clones at the output given by 
	\begin{equation}
		F_{i}= \braket{\psi|\hat{\rho}_{i}|\psi}.
	\end{equation}
	Here $\hat{\rho}_i$ is the reduced density matrix for a single clone, and $\ket{\psi}$ is the state to be copied. 
	For a symmetric cloner, this fidelity is identical for all the output clones.
	
	The single-copy fidelity for a $d$-dimensional $N \xrightarrow[]{}M$ optimal cloner ($N$ initial copies producing $M$ clones) was derived to be \cite{werner1998optimal}
	
	\begin{equation}
		F=\frac{M-N + N(M+d)}{M(N+d)},
	\end{equation}
	which we will use to verify whether we achieve optimal cloning for the spatial degrees of freedom of light.
	
	\subsection{Single photon input}
	First we demonstrate optimal cloning for the simplest case: a $1\xrightarrow{}2$ optimal quantum cloner. The input state to be cloned is a target spatial mode denoted by index $t$,  
	\begin{equation}
		\ket{1_{t}}_s = \hat{a}_{t,s}^{\dag}\ket{0}_s.
	\end{equation}
	
	We employ  Laguerre-Gauss (LG) spatial modes as the target modes in order to highlight the optimal cloning of OAM photons. At $z=0$, a LG beam mode is given in cylindrical coordinates by $\text{LG}_{m,\ell}(r,\phi)= \text{exp}\left(-\frac{r^2}{2w_0^2} +i\ell\varphi  \right) \left( \frac{r}{w_0} \right)^{|\ell|} L_{m}^{|\ell|}\left( \frac{r^2}{w_0^2}\right) $. Here $m$ is the radial index of the beam's intensity rings, $\ell$ is the topological charge of the beam (which indicates the OAM), $w_0$ is the Gaussian waist radius, and $L_{m}^{|\ell|}\left( x\right)$ are the associated Laguerre polynomials. For simplicity, we will consider only single-ring modes ($m=0$) and a pump beam with LG mode $\ell=0$. The creation operator for the input target mode is therefore given by $\hat{a}_{t,s}^{\dag}= \int 
	\text{d}^2\textbf{q}_s \mathcal{LG}_{\ell=t}(\textbf{q}_s)\hat{a}^{\dag}(\textbf{q}_s)$.
	After the DFG process, the output state is given by Eq.~(\ref{ddout}), where our input signal mode is now the target mode $t$. 
	
	Let us consider a two-level system with only $\ell=\pm 1$ modes, which is realised by means of appropriate filters after the down-conversion process, with  LG mode $\ell=+1$ as target. Assuming down-conversion as well as filtering is successful and perfect phase-matching occurs,  the normalised output state reads
	\begin{equation}\label{eqn:normbiphoton}
		\ket{\psi_{\text{out}}}= \frac{1}{\sqrt{3}} (\sqrt{2} \ket{2_{1}}_s \ket{1_{-1}}_i +\ket{1_{1},1_{-1}}_s \ket{1_{1}}_i ).
	\end{equation}
	We determine the expansion coefficients $\mathcal{N}^0_{-1,+1}$ (Eq.~(\ref{ddout})) to be a constant in terms of the waist radius $w_0$ (See Supplementary Information (SI)), which is included in the description of the idler photon as seen in Eq.~(\ref{eqn:prodidler}).  
	The expectation value of the number of target photons in the output state is given by
	$\braket{\psi_{\text{out}}|\hat{a}_{1,s}^\dagger \hat{a}_{1,s}|\psi_{\text{out}}}= \frac{1}{3}(4+1)=\frac{5}{3}$.
	However, in order to recover the single photon fidelity, we need to divide the average number of target photons in the output state by the number of copies.  
	This leads to $F_{\text{cal}}=\tfrac{5}{3}\times\tfrac{1}{2}=\tfrac{5}{6}$, which is the exact fidelity for the 2-dimensional $1\xrightarrow[]{}2$ optimal quantum cloner. The down-conversion process produces, with high probability amplitudes, modes with lower OAM values.
	This leads to different expansion coefficients $\mathcal{N}^0_{-\ell,\ell}$ that depend on the azimuthal index $\ell$. To obtain the same single photon fidelity as for optimal $1\xrightarrow{}2$ $d$-dimensional cloners, we require the expansion coefficients for different $\ell$ values to be equal. In order to realise this flat OAM spectrum, a few methods have been suggested which include pump beam shaping \cite{liu2020increasing,bornman2021optimal}, Procrustean filtering \cite{dada2011experimental}, and using modified LG spatial modes \cite{chen2013making}, which we discuss in the SI. Provided that this can be achieved, for the $d$-dimensional case (where $d=2 \ell_{\text{max}}$), the normalised output state is given by
	\begin{equation*}
		\ket{\psi_{\text{out}}}=\tfrac{1}{\sqrt{2+(d-1)}} \left(\sqrt{2}\ket{2_{t}}_s \ket{1_{t^{*}}}_i +\sum_{\substack{j=-\tfrac{d}{2}\\ j \neq t }}^{\tfrac{d}{2}} \ket{1_{t}, 1_{j}}_s \ket{1_{j^*}}_i  \right),
	\end{equation*}
	
	which leads to the same single photon fidelity as for optimal $1\xrightarrow{}2$ $d$-dimensional cloners,
	$F=\frac{d+3}{2(d+1)}$.

	\subsection{Fock state input}
	We now consider the 2-level case where we have $n$ input signal photons in our target state ($\ell=+1$ LG mode). The normalised output state when a single pump photon is down-converted is given by
	\begin{equation}
		\begin{split}
			\ket{\psi_{\text{out}}}&= \tfrac{1}{\sqrt{n+2}} (\sqrt{n+1} \ket{(n+1)_{1}}_s \ket{1_{-1}}_i + \ket{n_{1},1_{-1}}_s \ket{1_{1}}_i) .\\
		\end{split}
	\end{equation}
	The single photon fidelity reads
	\begin{equation}
		F_{\text{cal}}= \frac{\braket{\psi_{\text{out}}|\hat{a}_{1,s}^\dagger \hat{a}_{1,s}|\psi_{\text{out}}}}{n+1}=\frac{n+ (n+1)^2}{(n+1)(n+2)}, 
	\end{equation}
	which is the 2-dimensional  $N\xrightarrow{}N+1$ optimal cloning fidelity.
	
	For the $N\xrightarrow{}M$ case, $q=M-N$ photons must be down-converted. The output state corresponds to the term in the expansion of the unitary operator where a fixed number of $q$ photons are down-converted,
	\begin{equation}
		\begin{split}
			\ket{\psi_{\text{out}}}&= \frac{1}{\mathcal{N}} \frac{1}{q!}(\hat{a}_{1,s}^{\dag}\ \hat{a}_{-1,i}^{\dag}+ \hat{a}_{-1,s}^{\dag}\hat{a}_{1,i}^{\dag})^q \ket{n_{1}}_s \ket{0}_i  \\
			&= \frac{1}{\mathcal{N}} \sum_{k=0}^{q}\sqrt{\binom{n+k}{n}}  \ \ket{(n+k)_1,(q-k)_{-1}}_s \\
			& \hspace{2.5cm} \otimes \ket{k_{-1},(q-k)_{1}}_i , \\
		\end{split}
	\end{equation}
	where $\mathcal{N}= \sum_{k=0}^{q}   \sqrt{\binom{n+k}{n}}$ is the normalisation factor. The single photon fidelity agrees with the optimal $N\xrightarrow{}M$ single photon fidelity for qubits:
	\begin{equation}
		\begin{split}
			F_{\text{cal}}&= \frac{\braket{\psi_{\text{out}}|\hat{a}_{1,s}^\dagger \hat{a}_{1,s}|\psi_{\text{out}}}}{m} = \frac{\sum_{k=0}^{m-n} \binom{n+k}{n} (n+k) }{m \sum_{k=0}^{m-n}  \binom{n+k}{n}} \\
			&=  \frac{m+n+mn}{m (n+2)} .\\
		\end{split}
	\end{equation}
	We lastly consider the most general case: a $d$-dimensional $N\xrightarrow{}M$ optimal cloner. The output state can be written as
	\begin{equation}
		\begin{split}
			\ket{\psi_{\text{out}}}&= \frac{1}{\mathcal{N}} \frac{1}{q!} \left(\sum_{\ell=1}^{\tfrac{d}{2}} \left(\hat{a}_{\ell,s}^{\dag}\ \hat{a}_{-\ell,i}^{\dag}+  \hat{a}_{-\ell,s}^{\dag}\hat{a}_{\ell,i}^{\dag}\right)\right)^q  \ket{n_{t}}_s \ket{0}_i \\
			&=\frac{1}{\mathcal{N}} \sum_{q_1+\dots+q_d=q} \sqrt{\binom{n+q_t}{n}} \\
			&\hspace{1cm} \ket{(n+q_t)_t,(q_{t+d/2})_{-t}}_s \ket{(q_t)_{-t},(q_{t+d/2})_{t}}_i \\
			&\hspace{0.7cm}\otimes  \prod_{\substack{\ell=1\\ \ell \neq t }}^{\tfrac{d}{2}}\ \ket{(q_\ell)_{\ell},(q_{\ell+d/2})_{-\ell},}_s \ket{(q_\ell)_{-\ell},(q_{\ell+d/2})_{\ell}}_i ,
		\end{split}
	\end{equation}
	where $\mathcal{N}=\sum_{q_1+\dots+q_d=q} \sqrt{\binom{n+q_t}{n}} $ is the normalisation factor. The summation is over all permutations that satisfy the condition $q_1+\dots+q_d=q$ and $q_t$ is the number of photons in the target mode. To evaluate the summation over all permutations, we can directly sum over the number of ways photons are distributed into the target (correct) and non-target (incorrect) spatial modes (see SI for a classical analogy). We consider $k$ signal photons that are in the incorrect spatial modes and therefore $q_t=q-k$. The number of ways that $k$ photons are distributed into $d-1$ incorrect spatial modes is given by the number of $k$-combinations with repetition $C^{d-1}_{k,\text{rep}}= \frac{(d-1 + k-1)!}{k!(d-1-1)!}=\binom{d+k-2}{k}$.
	We therefore obtain the single photon fidelity 
	\begin{equation}\label{eq:ddimfidcal}
		\begin{split}
			F_{\text{cal}}&=\frac{\sum_{k=0}^q \binom{n+(q-k)}{n}\binom{d+k-2}{k} (n+(q-k)) }{m \sum_{k=0}^q \binom{n+(q-k)}{n}\binom{d+k-2}{k}}\\
			&=\frac{m-n+n(m+d)}{m(n+d)} , \\
		\end{split}
	\end{equation}
	which is the optimal fidelity for $d$-dimensional quantum cloning. Therefore we have shown that stimulated parametric down-conversion achieves $N\xrightarrow{} M $ $d$-dimensional optimal quantum cloning.
	\\

	\section{Discussion and Conclusion}
	In this paper, we have first shown that the process of DFG can be described for single photons with spatial modes. This is done by considering a general mode decomposition of the mode function of the down-conversion process. This method also allows us to see that the output state due to DFG is a separable state. We show the advantage of such a formulation of DFG by demonstrating that StimPDC achieves $d$-dimensional optimal quantum cloning with the spatial modes of light, which we explicitly show for OAM light. 
\linebreak

	\onecolumngrid
	\setcounter{equation}{0}
	\renewcommand{\theequation}{A\arabic{equation}}

	\section*{Supplementary Information}
		\appendix
	\subsection*{Overlap coefficients $\mathcal{N}_{k,m}^{p}$}
	Following from Eq.~\eqref{deq:dfgco} of the main paper, we show the standard calculation of the basis expansion coefficients, 
	\begin{equation}
		\begin{split}
			\mathcal{N}_{k,m}^{p}&= \int \text{d}^2\textbf{q}_2 \text{d}^2\textbf{q}_3 \mathcal{I}_{k}^{*}(\textbf{q}_3) \mathcal{S}_{m}^{*} (\textbf{q}_2) \Phi(\textbf{q}_2,\textbf{q}_3) \\
			&=\frac{1}{(2\pi)^2} \int \text{d}^2\boldsymbol{\rho} \mathcal{I}_{k}^{*}(\boldsymbol{\rho}) \mathcal{S}_{m}^{*} (\boldsymbol{\rho}) \mathcal{U}_{p}(\boldsymbol{\rho}),
		\end{split}
	\end{equation}
	with normalised Laguerre-Gauss spatial modes at $z$=0:
	
	\begin{equation}
		\text{LG}_{p,\ell}(r,\varphi)=\sqrt{\frac{2^{|\ell|+1} p!}{\pi \Gamma(p+|\ell|+1)w_0^2}}\  \text{exp}\left(-\frac{r^2}{w_0^2}  \right) \left( \frac{r}{w_0} \right)^{|\ell|} L_{p}^{|\ell|}\left( \frac{2r^2}{w_0^2}\right) \text{exp}\left(i\ell\varphi  \right)  .
	\end{equation}
	Here $r$ and $\varphi$ represent the polar coordinates, $w_0$ is the waist radius of a Gaussian beam, $p$ is the radial index and $\ell$ is the topological charge of the beam. The gamma function can be expressed as $\Gamma(n)=(n-1)!$
	
	For a Gaussian pump beam, the overlap integral becomes
	\begin{equation}
		\begin{split}
			\mathcal{N}_{\substack{\ell_i,\ell_s\\ p_i,p_s\\}}^{0,0}=\frac{1}{(2\pi)^2} \int \text{d}^2\boldsymbol{\rho} \text{LG}_{p_i,\ell_i}^{*}(\boldsymbol{\rho}) \text{LG}_{p_s,\ell_s}^{*} (\boldsymbol{\rho}) \text{LG}_{0,0}(\boldsymbol{\rho}) .
		\end{split}
	\end{equation}
	
\begin{figure}
		\includegraphics[width=0.6\linewidth]{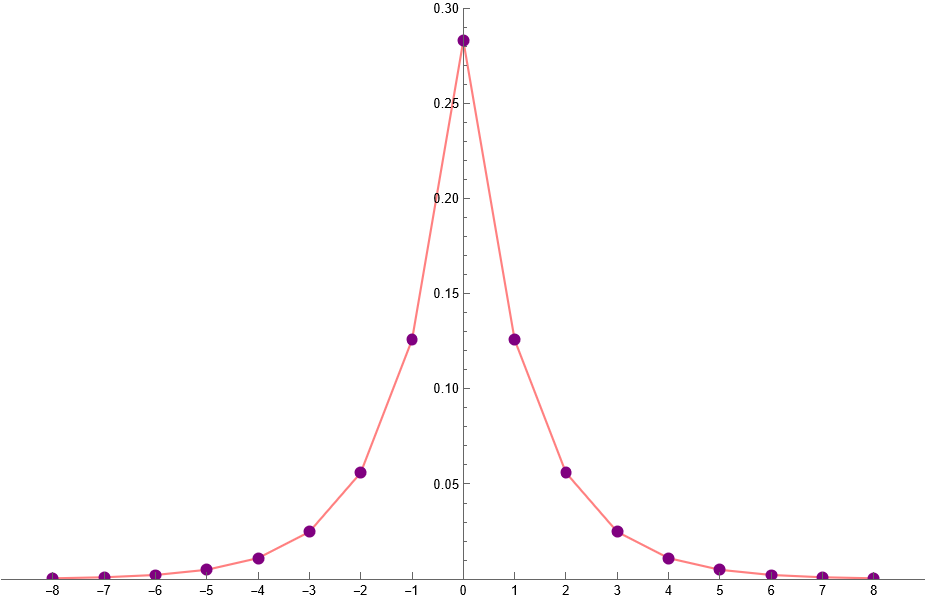}
		\centering
		\caption{\label{fig:sbw} Plot of $P_{\ell,-\ell}$ i.e. the OAM spectrum of down-conversion with $p$=0 and $w_0=1$.}
	\end{figure}

	We evaluate the overlap coefficients for the case where $w_p=w_i=w_s=w_0$. We also only consider single-ring beams where $p=0$.
	
	\begin{equation}
		\begin{split}
			\mathcal{N}_{\substack{\ell_i,\ell_s\\ 0,0\\}}^{0,0}&= \int \text{d}^2\boldsymbol{\rho} \text{LG}_{0,\ell_i}^{*}(\boldsymbol{\rho}) \text{LG}_{0,\ell_s}^{*} (\boldsymbol{\rho}) \text{LG}_{0,0}(\boldsymbol{\rho})\\
			&=  \int \text{d}^2 r\ r\text{d}^2\varphi \left(\sqrt{\frac{2^{|\ell_i|+1} }{\pi \Gamma(|\ell_i|+1)w_0^2}}\  \text{exp}\left(-\frac{r^2}{w_0^2}  \right) \left( \frac{r}{w_0} \right)^{|\ell_i|} \text{exp}\left(-i\ell_i \varphi  \right)\right)\\
			& \left(\sqrt{\frac{2^{|\ell_s|+1} }{\pi \Gamma(|\ell_s|+1)w_0^2}}\  \text{exp}\left(-\frac{r^2}{w_0^2}  \right) \left( \frac{r}{w_0} \right)^{|\ell_s|} \text{exp}\left(-i\ell_s \varphi  \right)\right)\left(\sqrt{\frac{1} {\pi w_0^2}}\  \text{exp}\left(-\frac{r^2}{ w_0^2}  \right) \right) \\
			&=\frac{1}{ w_0} \sqrt{\frac{2}{\pi}} \frac{2}{ 3}^{|\ell_s|+1} \\ 
		\end{split}
	\end{equation}
	We observe that the overlap coefficients for modes with the same value of topological charge $\pm \ell$ are equal. Therefore in the 2-dimensional case of optimal cloning, the overlap coefficients are equal, e.g. $\mathcal{N}_{+1,-1}^0=\mathcal{N}_{-1,+1}^0=\frac{1}{ w_0} \sqrt{\frac{2}{\pi}} \frac{4}{ 9} $ The OAM spectrum of the down-conversion process is obtained when considering the $\ell$ dependence of the overlap coefficients $P_{\ell,-\ell}=(\mathcal{N}_{\ell_{i},\ell_{s}}^{0,0})^2$, as shown in Fig.\ \ref{fig:sbw}. 
	\subsection*{Flat $\ell$ distribution}
	
	Utilising the OAM of light is a promising way to realise many $d$-dimensional quantum communication schemes. However, for schemes that utilise nonlinear optical processes, a challenge encountered is the limited OAM distribution of the nonlinear processes. In the following, we briefly review three methods suggested for overcoming this problem; employing ideal OAM beams, post-selection filtering, and pump beam shaping.

	We first consider ideal OAM beams. Here we make use of OAM beams that have a radius that is independent of the topological charge $\ell$. Two options are the perfect vortex beams and modified Laguerre Gauss (MLG) beams. 
	Here we consider the so-called modified Laguerre Gauss beams of the form\cite{chen2013making}:
	\begin{equation}
		MLG^{l,N}_{0}(r,\phi)=|LG^{l}_{0}(r,\phi)|\mathrm{exp}(i\ lN \phi)
	\end{equation}
	where $l$ is some fixed constant. These are beams with a fixed radial intensity given by $|LG^{l}_{0}(r,\phi)|$, while the total OAM is $\ell=lN$. When the azimuthal number that determines the width of the OAM ring is $l=\pm 1$, the MLG modes can be considered as analytic functions since 
	\begin{equation}
		r^l e^{ilN\phi}=(r e^{iN\phi})^l= (r(\mathrm{cos}(N\phi)+i\mathrm{sin}(N\phi)))^l=(x+iy)^l ,
	\end{equation}
	has complex derivatives. Thus propagation dynamics of MLG beams can be studied within the frame of standard LG beams with the following decomposition,
	\begin{equation}
		MLG= \sum_{p=0} c_p \text{LG}_{p}^{l N}. 
	\end{equation}
	
	The overlap of idler and signal photons with MLG modes with a Gaussian pump is calculated in the following:
	\begin{equation}\label{eqn:mlg}
		\begin{split}
			\mathcal{N}_{\substack{l_i M,l_s N\\ 0,0\\}}^{0,0}&= \int \text{d}^2\boldsymbol{\rho} \text{MLG}_{0,l_i M}^{*}(\boldsymbol{\rho}) \text{MLG}_{0,l_s N}^{*} (\boldsymbol{\rho}) \text{LG}_{0,0}(\boldsymbol{\rho})\\
			&=  \int \text{d}^2 r\ r\text{d}\varphi \left(\sqrt{\frac{2^{|l_i|+1} }{\pi \Gamma(|l_i|+1)w_0^2}}\  \text{exp}\left(-\frac{r^2}{w_0^2}  \right) \left( \frac{r}{w_0} \right)^{|l_i|} \text{exp}\left(-i l_i M \varphi  \right)\right)\\
			&\left(\sqrt{\frac{2^{|l_s|+1} }{\pi \Gamma(|l_s|+1)w_0^2}}\  \text{exp}\left(-\frac{r^2}{w_0^2}  \right) \left( \frac{r}{w_0} \right)^{|l_s|} \text{exp}\left(-i l_s N \varphi  \right)\right) \left(\sqrt{\frac{2} {\pi 4 w_0^2}}\  \text{exp}\left(-\frac{r^2}{4 w_0^2}  \right) \right)\\
			&=\sqrt{\frac{2^{|l_i|+|l_s|+3} }{4\pi^3 |l_i|! |l_s|! w_0^6}}\  \int \text{d}^2 r\ r \text{exp}\left(-\frac{9 r^2}{4 w_0^2}  \right) \left( \frac{r}{w_0} \right)^{|l_i|+|l_s|} \delta_{l_iM, l_sN}.
		\end{split}
	\end{equation}
	For fixed radial intensity, $l_i=l_s=l$,
	\begin{equation}
		\begin{split}
			\mathcal{N}_{\substack{l M,l N\\ 0,0\\}}^{0,0} &=\sqrt{\frac{2^{|l|+|l|+3} }{4\pi^3 |l|! |l|! w_0^6}}\  \int \text{d}^2 r\ r \text{exp}\left(-\frac{9 r^2}{4 w_0^2}  \right) \left( \frac{r}{w_0} \right)^{|l|+|l|} \delta_{lM, lN}\\
			&=\frac{2^{|l|}}{ w_0} \sqrt{\frac{2}{\pi}} \frac{4}{ 9}^{|l|+1} .
		\end{split}
	\end{equation}

	It is important that from the first line in Eq.~(\ref{eqn:mlg}), the product of the idler and signal spatial modes for a given basis must result in another spatial mode of that basis. Therefore when we consider LG and MLG beams, we consider single-ring beams where $p=0$. We are also restricted to cases where both values of signal and idler topological charge $\ell$ must satisfy $\ell \geq 0$ or at least one of the values must be zero.

	The second method of post-selection/Procrustean filtering (entanglement concentration) is widely employed \cite{dada2011experimental}. For such a filtering, we only project onto the subspace that ensures the mode amplitudes will be equal. In this case, we would need to project only onto the topological charge $\ell$ of the beam, regardless of the radial charge $p$, and more explicitly, we are restricted to the $p=0$ beams. However, this comes at the cost of a very reduced number of counts, due to the limited subspace that we project onto. 
	
	In order to get around the shortcomings associated with filtering, pump beam shaping was suggested as an alternative method. Here the input pump beam of the down-conversion process is shaped in order to achieve a flat and broad OAM spectrum. As demonstrated \cite{torres2003preparation,xu2022manipulating}, the input pump beam typically utilised is a superposition of OAM states.

	\subsection*{Optimal cloning fidelity}
	We can describe the cloning process that occurs in stimulated emission in terms of a classical example. In stimulated down-conversion, we have $n$ initial photons in our target mode and $q=m-n$ signal photons that are created by the down-conversion process. Each created signal photon has a certain probability to be in any of the $d$ considered spatial modes. As there are already initial photons in the target spatial mode, there is an increased chance for additional signal photons to join the target spatial mode due to Bose enhancement.  
	
	As this cloning process is probabilistic, we can formulate it in terms of distributing $q$ indistinguishable balls into $d$ bins. There is one ``correct" target bin and $d-1$ ``incorrect" bins that the balls can be distributed into. 
	
	We first consider the case where $k$ balls are distributed into $d-1$ incorrect bins while $q-k$ balls go into the correct bin. As each ball is indistinguishable, the ordering does not matter, so the number of ways to arrange the $k$ balls into the $d-1$ bins is given by the combination of $d-1$ bins taken $k$ at a time with repetition. This is considered with repetition as more than one ball can go into a specific bin.  Therefore the number of ways to distribute $k$ balls is given by $\tfrac{(d-1+k-1)!}{k!(d-1-1)!}= \binom{d+k-2}{k}$. This is the number of ways $k+d-2$ objects can be arranged divided by the number of ways $k$ and $d-2$ objects can be arranged. 
	We note here that this number is equivalent to the combination of $d-2$ bins taken $k$ times. E.g. for $\text{C}^n_r=\binom{n}{r}$, considering this combination with repetition gives $\text{C}^n_{r,rep}=\binom{n+r-1}{r}=\text{C}^{n+r-1}_r$. By allowing for repetition in a combination, we increase the size of the set ($n$) by the adding $r-1$ to it.
	
	If $q-k$ balls go into the correct bin, the total number of balls in the correct bin will be $n+(q-k)$. The number of ways to arrange $n+(q-k)$ balls when ordering matters is given by $(n+(q-k))!$. However, if the balls are indistinguishable we divide this expression by the number of ways to arrange the initial $n$ balls and the number of ways to arrange the additional $(q-k)$ balls. Therefore the number of ways the $q-k$ balls can be sorted into the correct bin is given by $\frac{(n+(q-k))!}{n!(q-k)!}=\binom{n+(q-k)}{n}$. 
	
	It now follows that as $k$ varies, the total number of ways that $q$ indistinguishable balls can be distributed into an already occupied correct bin and $d-1$ empty incorrect bins is $\sum_{k=0}^{q} \binom{n+(q-k)}{n} \binom{d+k-2}{k}$.

	The average number of balls in the correct bin is given by the sum over $k$ of the weighted distribution of balls in the correct bin. This is just the sum over $k$ of the number of balls in the correct bin ($n+(q-k)$) times its relative multiplicity divided by the total multiplicity,
	\begin{equation}
		\frac{\sum_{k=0}^{q} \binom{n+(q-k)}{n} \binom{d+k-2}{k} (n+(q-k))}{ \sum_{k=0}^{q} \binom{n+(q-k)}{n} \binom{d+k-2}{k}}.
	\end{equation}
	If we now divide this expression for the average number of balls in the correct bin by the total number of balls $n+q=m$, we would obtain the probability for a single ball to be in the correct bin. This corresponds to the single-clone fidelity for optimal symmetric $d$-dimensional quantum cloning (see Eq.~(\ref{eq:ddimfidcal}) in main paper).

\end{document}